**Pressurizing Field-Effect Transistors of Few-Layer $MoS_2$ in a Diamond Anvil Cell**

Yabin Chen[1], Feng Ke[2], Penghong Ci[1], Changhyun Ko[1], Taegyun Park[1], Sahar Saremi[1], Huili Liu[1,3], Yeonbae Lee[3], Joonki Suh[1], Lane W. Martin[1], Joel W. Ager[1,3], Bin Chen[2], Junqiao Wu[1,3,*]

[1]Department of Materials Science and Engineering, University of California, Berkeley, CA 94720, USA.

[2]Center for High Pressure Science and Technology Advanced Research, Shanghai 201203, China.

[3]Materials Sciences Division, Lawrence Berkeley National Laboratory, Berkeley, CA 94720, USA.

*Correspondence and requests for materials should be addressed to J.W. (wuj@berkeley.edu).

Y.C. and F.K. contributed equally to this work.

## ABSTRACT

Hydrostatic pressure applied using diamond anvil cells (DAC) has been widely explored to modulate physical properties of materials by tuning their lattice degree of freedom. Independently, electrical field is able to tune the electronic degree of freedom of functional materials via, for example, the field-effect transistor (FET) configuration. Combining these two orthogonal approaches would allow discovery of new physical properties and phases going beyond the known phase space. Such experiments are, however, technically challenging and have not been demonstrated. Herein, we report a feasible strategy to prepare and measure FETs in a DAC by lithographically patterning the nanodevices onto the diamond culet. Multiple-terminal FETs were fabricated in the DAC using few-layer $MoS_2$ and BN as the channel semiconductor and dielectric layer, respectively. It is found that the mobility, conductance, carrier concentration, and contact conductance of $MoS_2$ can all be significantly enhanced with pressure. We expect that the approach could enable unprecedented ways to explore new phases and properties of materials under coupled mechano-electrostatic modulation.

The key reason that semiconductors are extensively used in various device applications is that their physical properties can be conveniently tuned by external stimuli, such as electrical or optical fields, temperature, and stress/pressure.[1-6] By injecting excess charge carriers with electrostatic fields, electrical conductivity of a typical semiconductor could be dramatically modulated, and this effect lays the foundation for operation of the basic device unit of information technology, the field-effect transistor (FET). Electrostatic fields can also influence many other physical properties of materials. For example, doped molybdenum disulfide ($MoS_2$), as a representative two dimensional (2D) transition metal dichalcogenide, undergoes a sequence of semiconducting-metallic-superconducting phase transitions at high fields;[7, 8] and layered $TaS_2$ exhibits layer-dependent charge density waves that are also field-dependent.[9] On the other hand, high hydrostatic pressure, *i.e.*, isotropic pressure applied to a solid specimen via liquid or amorphous media using a diamond anvil cell (DAC), has been widely utilized to mechanically tune lattice parameters of solid materials, hence modulating their various order parameters, electronic structures, and interfacial coupling.[10-12] The hydrostatic pressure approach has the advantages of being non-intrusive, reversible, "clean" from contamination (unlike chemical doping), and able to impose high strain without cracking the sample (unlike uniaxial stress). For example, $MoS_2$ displays an isostructural transition from $2H_c$ to $2H_a$ at a pressure of ~20 GPa,[13, 14] and further, becomes a superconductor at 90 GPa with a transition temperature onset of 11.5 K.[15] However, the two modulations with electrical fields and hydrostatic pressure have been, thus far, totally separate; namely, existing experiments probe material response by either varying electrical fields at ambient pressure, or varying pressure at zero gating electrical field. Simultaneous application and tuning of pressure and fields onto a solid material have not been demonstrated. In this work, we report for the first time the simultaneous mechano-electrostatic tuning by investigating the behavior of few-layer $MoS_2$ FET nanodevices inside a DAC. It is found that the electrical performance (*i.e.*, carrier concentration, mobility, and contact quality) of the $MoS_2$ FETs can be significantly modulated with hydrostatic pressure. We expect that the strategy of fabricating FETs inside the DAC could pave the way to discovering and probing novel properties of solid materials.

Although plenty of electrical conductance measurements in a DAC have been demonstrated on bulk materials,[16-19] the fabrication and measurements of FET nanodevices inside a DAC are still lacking. This lack of studies is due to several immense technical challenges, particularly: 1. Electrical connection between the nanoscale semiconductor channel (*e.g.*, 2D semiconductors) and metal electrodes inside the DAC. Making such connections reliably and stably conductive under high pressure is already notoriously challenging for bulk materials. It becomes even more difficult for 2D materials owing to their intrinsically poor electrical contact with most metals, as well as the fact that the popularly used metal foil (Au or Pt) or silver paste can hardly be applied to the 2D flakes that have a very limited lateral size.[16] 2. Reliability of the dielectric layer. The routinely used, deposited $Al_2O_3$ or $HfO_2$ films can easily fracture under high pressure, resulting in considerable leakage currents. 3. Unlike bulk crystals or deposited thin films, it is difficult to transfer 2D flakes onto the desired microscale area on the diamond surface without degrading its crystal structure and physical properties.

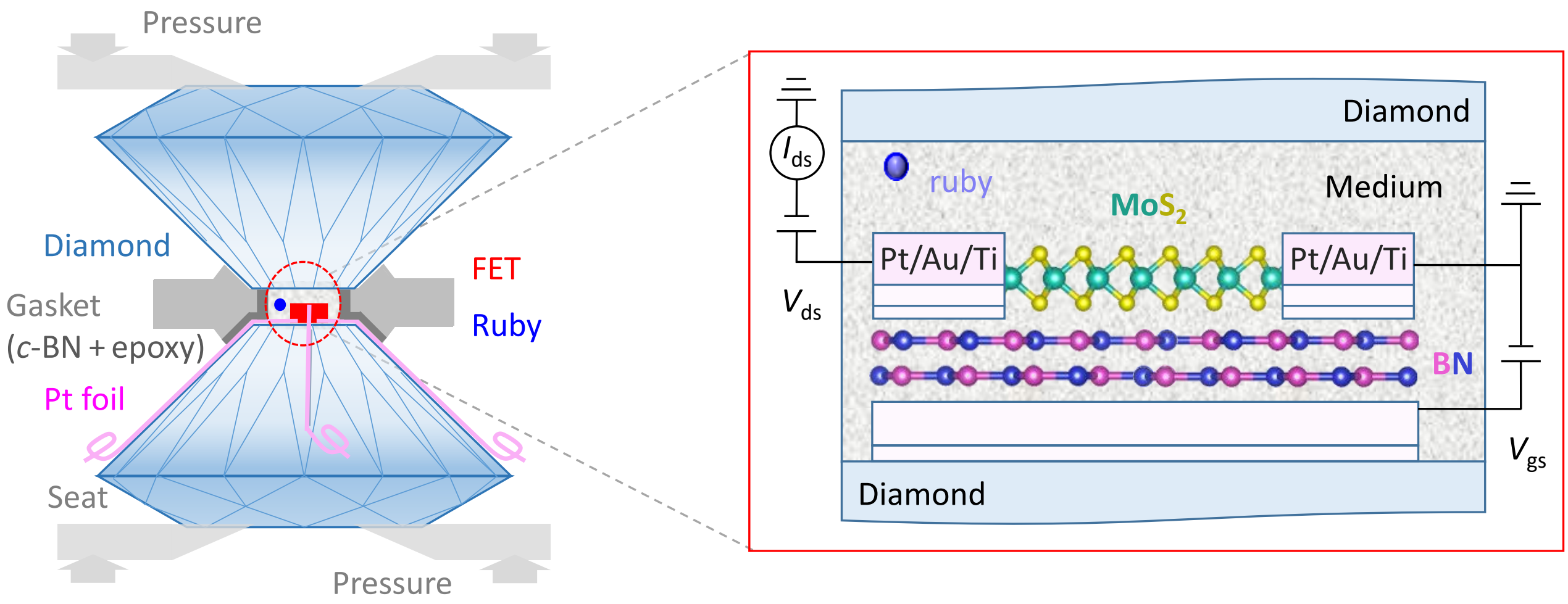

**Figure 1.** A FET nanodevice in a DAC. In this schematic, the layered $MoS_2$ and *h*-BN were used as the channel semiconductor and dielectric material, respectively. Each of the electrodes is composed of two parts, a patterned and deposited thin metal line (Au/Ti) inside the DAC chamber connected to a strip of Pt foil outside the chamber. The pressure was calibrated by ruby particles.

To achieve the operation of FET nanodevices under high pressures, we developed a strategy to directly fabricate multiple electrodes onto a 2D semiconductor flake and a layered dielectric inside the tiny DAC chamber. The multiple-electrode system is constituted of metal lines patterned by lithography and deposition onto the diamond surface (Supporting Information, Figure S1), which are connected to strips of Pt foil leading outside the chamber. We used hexagonal-boron nitride (*h*-BN) as the dielectric material due to its high breakdown field as well as layered structure that allows for good thickness control and compatibility with high pressure. Figure 1 depicts the configuration of a bottom-gated FET device in DAC. Briefly, cubic-BN (*c*-BN) was utilized as the coating to the steel gaskets to ensure electrical isolation between different electrodes.[20] Inside the gasket chamber, a strip of Au/Ti layer (50/5 nm) was first deposited as the gate electrode onto the diamond culet following photolithography patterning. The dielectric *h*-BN and channel semiconductor (few-layer $MoS_2$ in this work) were sequentially transferred onto the gate electrode. Afterwards, four more electrodes were further patterned and deposited onto the four corners of the $MoS_2$ flake for channel conductance and filed-effect measurements. Finally, each of these patterned electrodes was connected to a narrow strip of Pt foil (~5 μm thick) which was ultimately connected to the external electronics. Importantly, the Pt strips are required to extend into the pressure chamber across the diamond culet edge, because the patterned electrodes can rarely survive under high pressures once the gasket is sealed over the diamonds. After careful and optimal alignments, such nanodevices are robust and reproducible enough when the pressure exceeds ~50 GPa (=500,000 atmosphere), which is sufficiently high to drive structural transitions of most crystalline materials. The device fabrication yields a high success rate of ~80%.

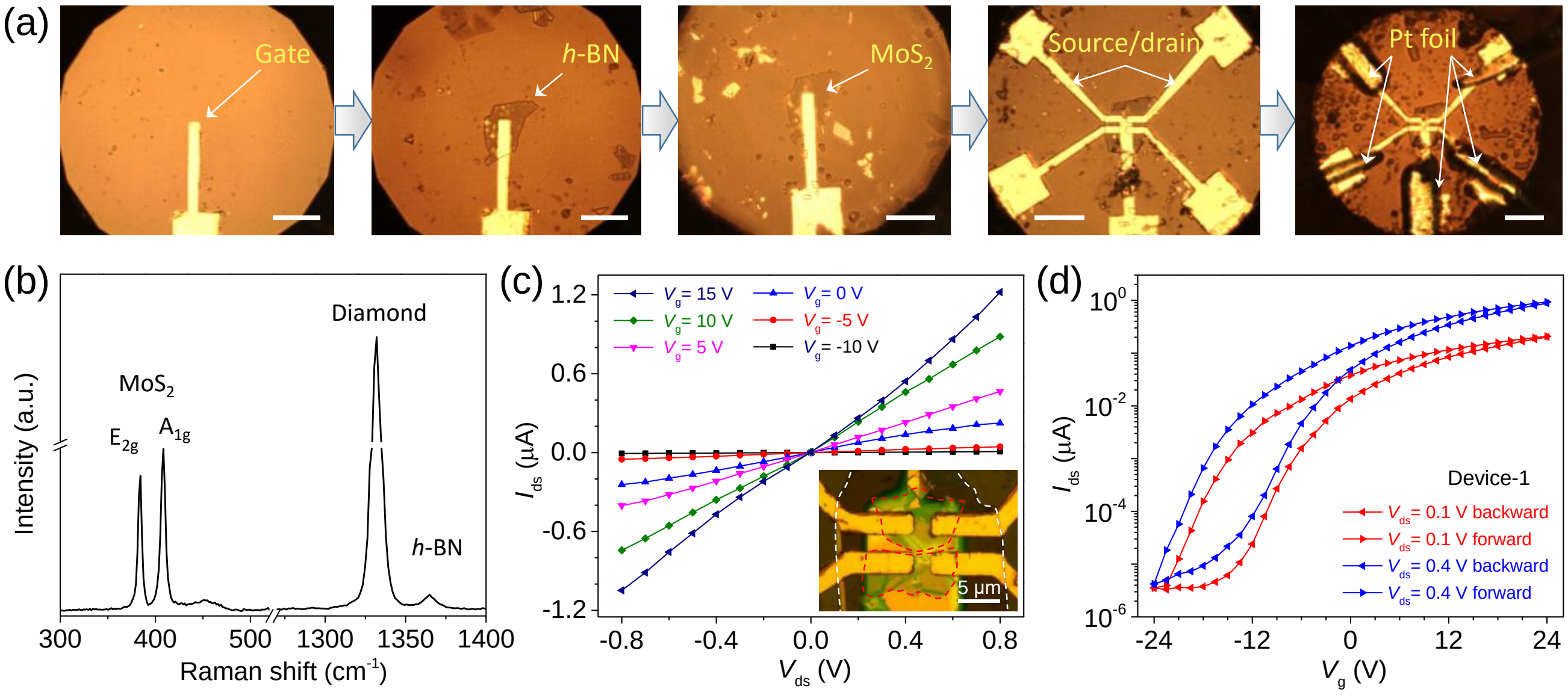

**Figure 2.** A FET of a few-layer $MoS_2$ on the diamond surface. (a) Fabrication process of a five-terminal $MoS_2$-FET on the diamond surface. The culet size is 300 μm. The scale bars are 50 μm. (b) Raman spectra of the transferred $MoS_2$ and *h*-BN flakes overlapped on the diamond surface. (c) Output curves of the $MoS_2$-FET nanodevice at ambient pressure. The gate voltage varies from -10 to 15 V. Inset is an optical image of the final device. The red and while dashed lines show the boundary of the $MoS_2$ and *h*-BN flakes, respectively. The 3.0 nm-thick $MoS_2$ flake in this device broke into two discontinued pieces during the exfoliation. (d) Transfer curves of the $MoS_2$-FET nanodevice under different source-drain voltages.

Following this method, few-layer $MoS_2$ FETs were successfully fabricated and measured. An actual process was recorded and shown in Figure 2a. The thickness of the transferred *h*-BN and $MoS_2$ flakes were ~91 nm and 3.0 nm, respectively, as confirmed by atomic force microscopy (AFM). It is noted that the gate electrode needs to be well positioned at the center of the pressure chamber, and the lateral size of the active region of the FET (namely, the channel and the gate materials) should be well within the radius of the gasket hole, otherwise the possible shrinkage of the gasket hole under high pressure may destroy the device. The Pt foil strips were manually aligned with the patterned electrodes under an optical microscope. Raman measurements were carried out on the overlapped $MoS_2$/*h*-BN region to confirm the successful transfer. As shown in Figure 2b, the four Raman peaks at 383.8, 408.1, 1331.8 and 1365.1 $cm^{-1}$ are consistent with the Raman modes of in-plane $E_{2g}$ of $MoS_2$, out-of-plane $A_{1g}$ of $MoS_2$, first-order $F_{2g}$ of diamond, and first-order $E_{2g}$ of *h*-BN, respectively (Supporting Information, Figure S2-S3).[21, 22] Figure 2c displays the obtained output characteristics between a pair of source-drain electrodes under various gate voltages (the other pair showed similar results). The channel current becomes significantly higher as the gate voltage increases toward positive values, consistent with the expected n-type property of exfoliated $MoS_2$.[23] The nearly linear *I-V* curves indicate reasonably good contact between the $MoS_2$ and the electrodes. Additionally, the transfer curves in Figure 2d prove excellent dielectric insulation of the *h*-BN flake. The current on/off ratio, $I_{on}/I_{off}$, is higher than $10^5$. The field effect mobility was extracted from the linear region of the conductance curves. The mobility for this device is ~ 4 $cm^2V^{-1}s^{-1}$ calculated using the equation $\mu = [dI_{ds}/dV_g] \times [L/(C_i W V_{ds})]$, where $L$ and $W$ are the channel length and width, respectively, $C_i$ is the capacitance per unit area between the channel and the gate electrode.[24] This mobility value is underestimated due to the non-negligible contact resistance. In a latter section, we will eliminate the effect of the contact resistance using four-probe van der Pauw method. The hysteresis in Figure 2d between forward and backward measurements may be attributed to unintentional doping effects from ambient water or oxygen, as reported before.[25] All of these results validate our method of device fabrication and measurements on the diamond culet.

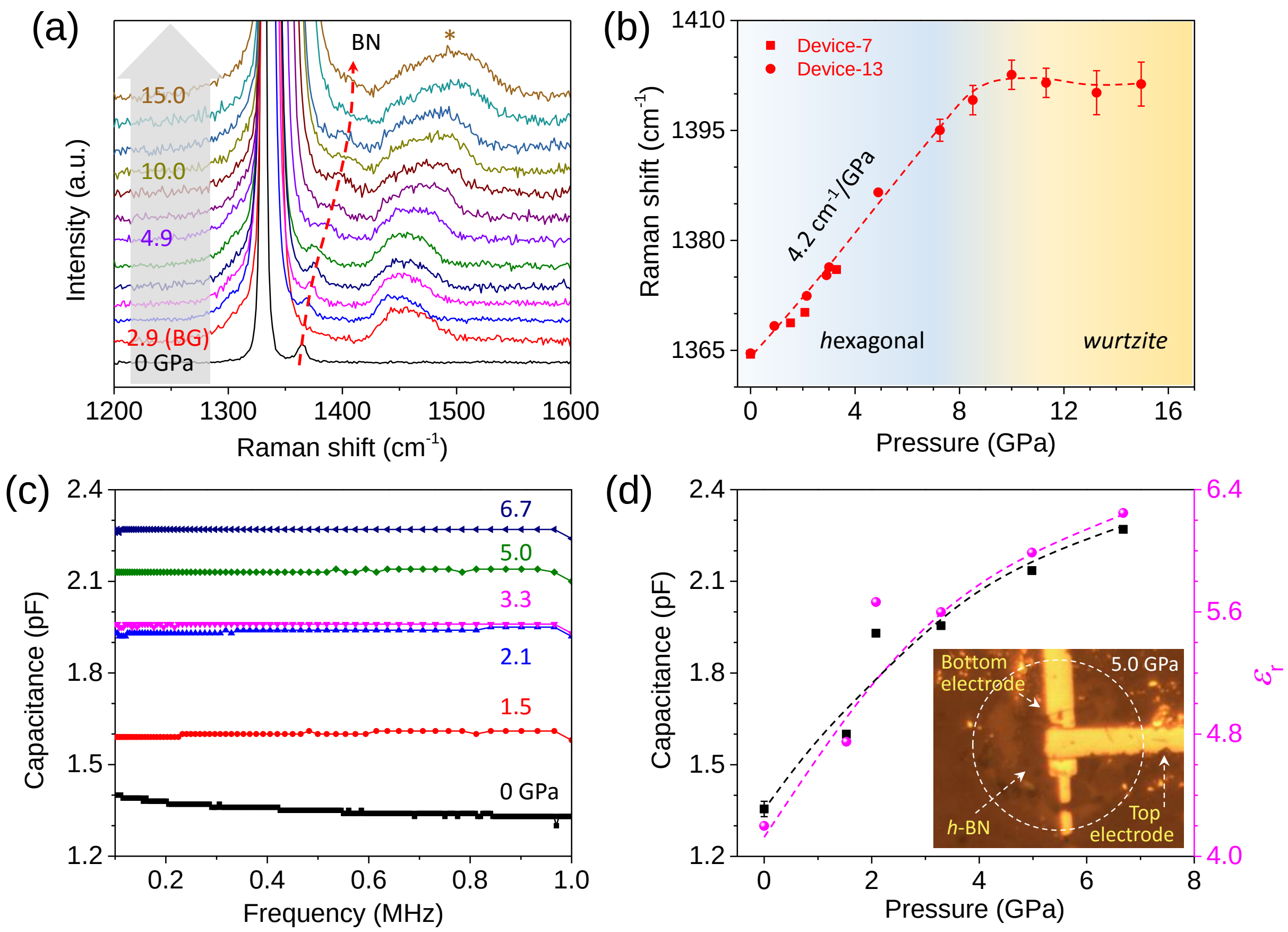

**Figure 3.** Pressure behavior of the dielectric layer, *h*-BN. (a) Raman spectrum of a 47 nm-thick *h*-BN under pressure. The red curve of 2.9 GPa was taken on the blank diamond surface as the background. The black curve of 0 GPa was taken before clamping the cell. The asterisk indicates a broad band arising from the pressure medium. (b) Pressure-dependent Raman shift of *h*-BN. The change in slope at ~ 9 GPa signifies a phase transition from hexagonal to wurtzite structure. The circle and square symbols represent two different *h*-BN flakes. (c) Capacitance measurements of *h*-BN in the DAC at different pressures. (d) Pressure-dependent capacitance and relative dielectric constant of *h*-BN. Inset is a typical *h*-BN device for the capacitance measurements imaged at 5.0 GPa, showing the bottom and top electrodes.

As the dielectric property of the *h*-BN may vary with pressure, its pressure behavior must be calibrated before we use it to analyze the FET data. We hence investigated the dielectric constant of *h*-BN under high pressure. Figure 3a shows the Raman spectrum of *h*-BN in the DAC up to a pressure of 15 GPa. The Raman peak of both *h*-BN and diamond exhibits a clear blue shift due to the pressure-induced lattice stiffening. By comparison to the background signal, the board band at ~1450 $cm^{-1}$ comes from the pressure medium. As shown in Figure 3b, the Raman shift of *h*-BN first increases monotonically and then becomes flat, implying a phase transition from the hexagonal phase to the wurtzite phase. The stiffening rate of the hexagonal phase Raman is 4.2 $cm^{-1}$/GPa, which is in agreement with literature.[22] More importantly, we carried out capacitance measurements of the *h*-BN layer in the DAC in order to determine its change in dielectric constant under high pressure, as shown in Figure 3c and 3d. The extracted capacitance increases monotonically from 1.3 pF at 0 GPa to 2.3 pF at 6.7 GPa. Its relative dielectric constant was calculated using the expression $C = A \times C_i = A \times (\varepsilon_0 \varepsilon_r / t)$, where $A$ is the overlap area between the *h*-BN flake and electrodes, $\varepsilon_r$ and $t$ are the relative dielectric constant and layer thickness, respectively, and $\varepsilon_0 = 8.85 \times 10^{-12}$ F $m^{-1}$ is the vacuum dielectric constant.[26] The reduction in thickness of *h*-BN as a function of pressure was calculated using the existing data reported based on x-ray diffraction.[27] The obtained $\varepsilon_r$ becomes 1.5 times higher at 6.7 GPa than that at ambient pressure, an effect that can be explained by the pressure-induced bandgap ($E_g$) reduction of *h*-BN [28] following the empirical Moss relation, $\varepsilon_r^2 E_g \approx$ 95 eV for most semiconductors.[29] As its bandgap becomes smaller, *h*-BN has a higher $\varepsilon_r$ and can more effectively screen the gating fields for the channel $MoS_2$.

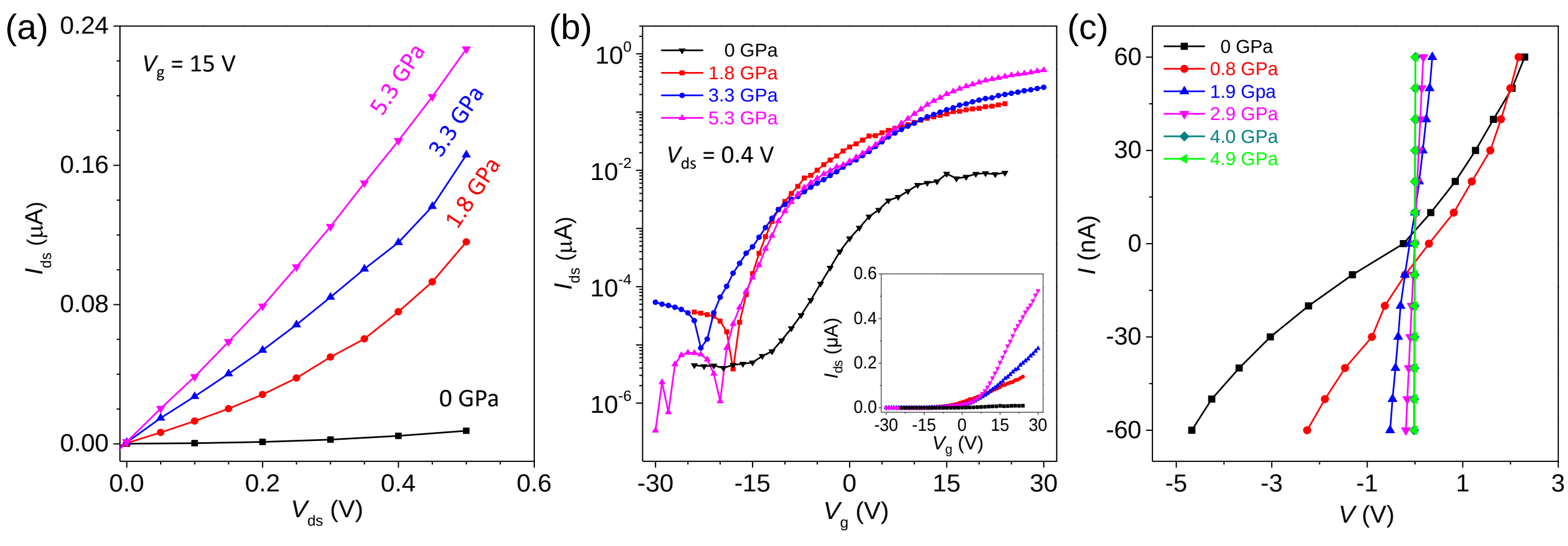

**Figure 4.** Transport measurements of a $MoS_2$ FET in the DAC. (a) Output curves of a bilayer-$MoS_2$

FET under different pressures. (b) Transfer curves of the same bilayer-$MoS_2$ FET under different pressures. Inset is the same data shown in linear scale. (c) Four-probe *I*-*V* curves of a $MoS_2$ FET obtained through the van der Pauw method in the DAC, where the $MoS_2$ flake was 3-5 layers.

Pressure-modulated transport characteristics of $MoS_2$ FETs were investigated and a typical one is shown in Figure 4. For a given $V_g$, the source-drain current $I_{ds}$ increases dramatically as the pressure increases, which is in agreement with the expected pressure-induced reduction in bandgap and resistance of $MoS_2$.[13] Figure 4b displays the pressure-dependent transfer characteristics of a $MoS_2$ FET. The on-state current clearly increases as the pressure increases from 0 to 5.3 GPa. We further measured the real channel resistance, $R_{ch}$, using the four-probe van der Pauw method on the $MoS_2$ flake. In this method, a constant current of 100 nA was flowed between two adjacent electrodes, and the other, opposite pair of electrodes probed the voltage drop. As seen from Figure 4c, the $MoS_2$ channel becomes monotonically more conductive as pressure increases. The *I*-*V* curve also becomes more linear at higher pressures, indicating that the pressure improves the electrical contact. The decreased $R_{ch}$ contributed significantly to the improved two-probe conductance in Figure 4a and 4b.

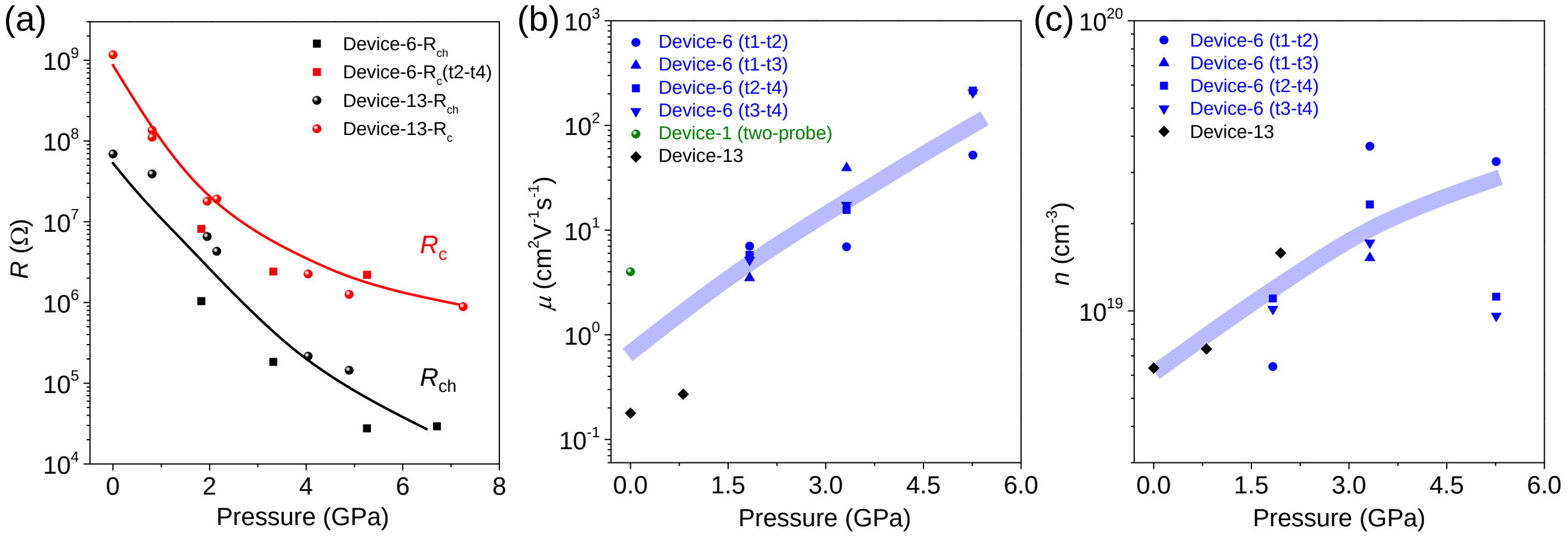

**Figure 5.** (a) Channel resistance (black) and contact resistance (red) of $MoS_2$-FET devices as a function of pressure. (b) Pressure-dependent mobility of $MoS_2$ flakes in the DAC. The t1 to t4

indicate the electrode number from the same FET nanodevice (further details in Supporting Information, Figure S4). (c) Extracted carrier concentration of $MoS_2$ flakes as a function of pressure. The $MoS_2$ flakes are bilayer and 3-5 layers in device 6 and 13, respectively. All lines are drawn to guide the eye.

Next, we separate the contact resistance $R_c$ from the channel resistance $R_{ch}$ using the four-probe data in Figure 4c via $R_{tot} = R_{ch} + 2R_c$, where $R_{tot}$ is the obtained two-probe total resistance. $R_{ch}$ is extracted by the van der Pauw method (More details in Supporting Information). The resultant data is plotted in Figure 5a and shows two effects: 1. the contact resistance $R_c$ drops drastically with pressure, and more importantly, 2. the channel resistance $R_{ch}$ also decreases rapidly with pressure. The improved contact quality is expected from the reduced Schottky barrier at the contact. The Schottky barrier height between n-type $MoS_2$ and Ti metal is reported to be ~50 meV, after a pinning of the Ti Fermi level near the conduction band of $MoS_2$ (the work function of Ti is ~4.3 eV, and electron affinity of $MoS_2$ is ~4.0 eV).[30] Theoretical calculations predict that hydrostatic pressure rapidly pushes down the conduction band minimum of few-layer $MoS_2$ toward a better alignment with the Fermi level of Ti.[31] As a result, the Schottky barrier height between the Ti and $MoS_2$ is reduced, resulting in a lower contact resistance. The high pressure may also mechanically push the electrode against the $MoS_2$ flake, effectively enlarging the contact area at the microscale, which contributes to lowering the contact resistance.

Following the analysis, we subsequently calculated the pressure dependence of the field-effect mobility $\mu$ and electron concentration $n$ of $MoS_2$, as shown in Figure 5b and 5c. The mobility was extracted from the transfer curves in Figure 4b using $\mu = [dI_{ds}/dV_g] \times [L/(C_i W V_{ds(ch)})]$, where $V_{ds(ch)}$ is the calibrated voltage drop solely onto the channel that eliminates the effects of contact resistance (more details in Supporting Information). The obtained mobility shows a significant improvement with pressure. This improvement can be, at least partially, attributed to the reduced electron effective mass $m^*$ under pressure that has been predicted by *ab initio* calculations.[32] According to the Drude model, $\mu$ is inversely proportional to $m^*$ as $\mu = e\tau/m^*$. Assuming that the relaxation time $\tau$ is less sensitive to external pressure (because pressure does not severely alter the electron-scattering centers in the channel such as impurity and surface condition of $MoS_2$), an improved $\mu$ is expected. Finally, the carrier concentration $n$ increases following $n = 1/(e\mu\rho_{ch})$, where $\rho_{ch}$ is the channel resistivity. This is consistent with the reduction in $m^*$ and in bandgap of $MoS_2$ under high pressure.[32] Generally, the unintentionally doped free electrons in exfoliated $MoS_2$ flakes originate from native defects such as sulphur vacancies. These native defects act as shallow donors and donate free electrons to the conduction bands of $MoS_2$ upon thermal excitation at room temperature. According to the Hydrogen defects model, the ionization energy of these shallow donors is 13.6 eV × $m^*/\varepsilon_r^2$, where $\varepsilon_r$ is the relative dielectric constant of $MoS_2$, which is expected to also follow the empirical Moss relation with its bandgap as discussed earlier for *h*-BN. The pressure driven reduction in both $m^*$ and bandgap of $MoS_2$ thus reduces the ionization energy of these shallow donors, resulting in an increase in free electron density. The degenerately high free electron density might also more strongly screen some of the carrier scattering centers, which helps to increase the electron mobility.

In conclusion, we report the behavior of few-layer $MoS_2$ flakes under both hydrostatic pressure and gating fields by assembling them into the FET configuration inside a DAC. For the first time, our technique enables simultaneous, *in-situ* electronic and structural modulation of the channel material by electrostatic and pressure means. It is found that the carrier mobility and density in the channel, as well as the contact conductance, can all be remarkably improved with the hydrostatic pressure. We believe that our methodology can be widely employed for unprecedented exploration of condensed matter under simultaneous tuning with electrical fields (electrostatic gating) and mechanical (pressure) forces.

## Methods

*Device preparation and characterization.*

The $MoS_2$ and *h*-BN flakes were mechanically exfoliated onto a Si substrate (300 nm oxide layer). They were sequentially transferred onto the gate electrode by a polydimethylsiloxane transfer method.[12] The flake thickness was confirmed by tapping-mode scanning of AFM (Nanoscope 3D). The micro-Raman measurements were carried out using the commercial Renishaw Raman spectroscope, and the laser wavelength was 488 nm. All devices were prepared by following the standard photolithography and lift-off process. The Ti/Au films as the inner electrodes were deposited by electron beam evaporation in a chamber with pressure lower than $10^{-7}$ Torr. The capacitance of the *h*-BN devices was measured by a LCR meter (Aligent E4980A).

*High pressure measurement in the DAC.*

Hydrostatic pressure was generated by a symmetric DAC with 300 μm culet size (type IA diamond). A mixture of *c*-BN powder and epoxy was used as the insulating coating for the steel gaskets. The fabrication method has been reported elsewhere.[20] The pre-indented thickness of gasket was within 40 μm, and a hole of 100 μm drilled by laser was utilized as the pressure chamber. Daphene 7373[33] was used as the pressure-transmitting medium. The thickness of the Pt foil was around 5 μm, and it was manually cut by a sharp blade into narrow strips (< 10 μm) as the outer electrodes. The pressure was calibrated by the photoluminescence of ruby particles placed next to the sample inside the pressure chamber.[34]

## Supporting Information

The Supporting Information is available free of charge on the ACS Publications website.

## Corresponding Author

*E-mail: wuj@berkeley.edu

## Notes

The authors declare no competing financial interest.

## Acknowledgments

This work was supported by the National Science Foundation under grant No. DMR-1306601. The device fabrication used facilities at the Lawrence Berkeley National Laboratory, which is supported by the Office of Science, Office of Basic Energy Sciences, of the U.S. Department of Energy under Contract No. DE-AC02-05CH11231. J.W., J.A., and Y.C. acknowledge support from the Singapore-Berkeley Research Initiative for Sustainable Energy (SinBeRISE). S.S. acknowledges support from the National Science Foundation under grant CMMI-1434147. L.W.M. acknowledges support from the Army Research Office under grant W911NF-14-1-0104. The laser milling setup in Advanced Light Source was supported by COMPRES (Grant No. EAR 11- 57758). Y.C. and F.K. gratefully thank Dr. Jinyuan Yan from Advanced Light Source for the help on gasket fabrication.